\documentclass[11pt]{article}
\usepackage{graphicx}
\usepackage{dcolumn}     
\usepackage{bm}

\newcommand{\BABARPubYear}    {04}
\newcommand{\BABARConfNumber} {163}
\newcommand{\SLACPubNumber} {10840}

\input babarsym

\usepackage{relsize}
\def\babar{\mbox{\slshape B\kern-0.1em{\smaller A}\kern-0.1em
    B\kern-0.1em{\smaller A\kern-0.2em R}}}

\def\en         {\ensuremath{e^-}\xspace}   
\def\ep         {\ensuremath{e^+}\xspace}
 
\def\epem       {\ensuremath{e^+e^-}\xspace}

\def\mup        {\ensuremath{\mu^+}\xspace}
\def\mun        {\ensuremath{\mu^-}\xspace} 
\def\mumu       {\ensuremath{\mu^+\mu^-}\xspace}

\def\taup       {\ensuremath{\tau^+}\xspace}
\def\taum       {\ensuremath{\tau^-}\xspace}

\def\ellell     {\ensuremath{\ell^+ \ell^-}\xspace}

\def\nub        {\ensuremath{\overline{\nu}}\xspace}
\def\nunub      {\ensuremath{\nu{\overline{\nu}}}\xspace}
\def\nub        {\ensuremath{\overline{\nu}}\xspace}
\def\nunub      {\ensuremath{\nu{\overline{\nu}}}\xspace}
\def\nue        {\ensuremath{\nu_e}\xspace}
\def\nueb       {\ensuremath{\nub_e}\xspace}

\def\num        {\ensuremath{\nu_\mu}\xspace}
\def\numb       {\ensuremath{\nub_\mu}\xspace}

\def\nut        {\ensuremath{\nu_\tau}\xspace}
\def\nutb       {\ensuremath{\nub_\tau}\xspace}

\def\nul        {\ensuremath{\nu_\ell}\xspace}
\def\nulb       {\ensuremath{\nub_\ell}\xspace}

\def\g     {\ensuremath{\gamma}\xspace}
\def\gaga  {\ensuremath{\gamma\gamma}\xspace}  
\def\ggstar{\ensuremath{\gamma\gamma^*}\xspace}

\def\piz   {\ensuremath{\pi^0}\xspace}

\def\ppz   {\ensuremath{\pi^0\pi^0}\xspace}
\def\pip   {\ensuremath{\pi^+}\xspace}
\def\pim   {\ensuremath{\pi^-}\xspace}
\def\pipi  {\ensuremath{\pi^+\pi^-}\xspace}

\def\pimp  {\ensuremath{\pi^\mp}\xspace}

\def\kaon  {\ensuremath{K}\xspace}
\def\Kbar  {\kern 0.2em\overline{\kern -0.2em K}{}\xspace}

\def\Kz    {\ensuremath{K^0}\xspace}
\def\Kzb   {\ensuremath{\Kbar^0}\xspace}
\def\KzKzb {\ensuremath{\Kz \kern -0.16em \Kzb}\xspace}
\def\Kp    {\ensuremath{K^+}\xspace}
\def\Km    {\ensuremath{K^-}\xspace}
\def\Kpm   {\ensuremath{K^\pm}\xspace}

\def\KpKm  {\ensuremath{\Kp \kern -0.16em \Km}\xspace}
\def\KS    {\ensuremath{K^0_{\scriptscriptstyle S}}\xspace} 
\def\KL    {\ensuremath{K^0_{\scriptscriptstyle L}}\xspace} 
\def\Kstarz  {\ensuremath{K^{*0}}\xspace}

\def\Kstar   {\ensuremath{K^*}\xspace}

\def\Kstarp  {\ensuremath{K^{*+}}\xspace}

\def\Dbar    {\kern 0.2em\overline{\kern -0.2em D}{}\xspace}

\def\Dz      {\ensuremath{D^0}\xspace}
\def\Dzb     {\ensuremath{\Dbar^0}\xspace}
\def\DzDzb   {\ensuremath{\Dz {\kern -0.16em \Dzb}}\xspace}
\def\Dp      {\ensuremath{D^+}\xspace}
\def\Dm      {\ensuremath{D^-}\xspace}

\def\DpDm    {\ensuremath{\Dp {\kern -0.16em \Dm}}\xspace}
\def\Dstar   {\ensuremath{D^*}\xspace}

\def\Dstarz  {\ensuremath{D^{*0}}\xspace}

\def\Dstarp  {\ensuremath{D^{*+}}\xspace}
\def\Dstarm  {\ensuremath{D^{*-}}\xspace}

\def\B       {\ensuremath{B}\xspace}
\def\Bbar    {\kern 0.18em\overline{\kern -0.18em B}{}\xspace}
\def\Bb      {\ensuremath{\Bbar}\xspace}
\def\BB      {\ensuremath{\B {\kern -0.16em \Bb}}\xspace}
\def\Bz      {\ensuremath{B^0}\xspace}
\def\Bzb     {\ensuremath{\Bbar^0}\xspace}
\def\BzBzb   {\ensuremath{\Bz {\kern -0.16em \Bzb}}\xspace}
\def\NBB     {\ensuremath{N_{\BB}}}

\def\Bu      {\ensuremath{B^+}\xspace}
\def\Bub     {\ensuremath{B^-}\xspace}

\def\Bpm     {\ensuremath{B^\pm}\xspace}

\def\BpBm    {\ensuremath{\Bu {\kern -0.16em \Bub}}\xspace}

\def\BorBbar    {\kern 0.18em\optbar{\kern -0.18em B}{}\xspace}
\def\DorDbar    {\kern 0.18em\optbar{\kern -0.18em D}{}\xspace}
\def\KorKbar    {\kern 0.18em\optbar{\kern -0.18em K}{}\xspace}

\def\jpsi     {\ensuremath{{J\mskip -3mu/\mskip -2mu\psi\mskip 2mu}}\xspace}

\mathchardef\Upsilon="7107
\def\Y#1S{\ensuremath{\Upsilon{(#1S)}}\xspace}

\def\FourS {\Y4S}

\mathchardef\Deltares="7101
\mathchardef\Xi="7104
\mathchardef\Lambda="7103
\mathchardef\Sigma="7106
\mathchardef\Omega="710A

\def\Deltabar{\kern 0.25em\overline{\kern -0.25em \Deltares}{}\xspace}
\def\Lbar{\kern 0.2em\overline{\kern -0.2em\Lambda\kern 0.05em}\kern-0.05em{}\xspace}
\def\Sigbar{\kern 0.2em\overline{\kern -0.2em \Sigma}{}\xspace}
\def\Xibar{\kern 0.2em\overline{\kern -0.2em \Xi}{}\xspace}
\def\Obar{\kern 0.2em\overline{\kern -0.2em \Omega}{}\xspace}
\def\Nbar{\kern 0.2em\overline{\kern -0.2em N}{}\xspace}
\def\Xb{\kern 0.2em\overline{\kern -0.2em X}{}\xspace}

\def\X {\ensuremath{X}\xspace}

\def\BR         {{\ensuremath{\cal B}\xspace}}

\def\bpsikl     {\ensuremath{\Bz \to \jpsi \KL}\xspace}

\def\Bzbtox     {\ensuremath{\Bzb \to \X}\xspace}

\def\invfb   {\ensuremath{\mbox{\,fb}^{-1}}\xspace}

\def\mus  {\ensuremath{\rm \,\mus}\xspace}

\def\mus        {\ensuremath{\,\mu{\rm s}}\xspace}    

\def\degrees{\ensuremath{^{\circ}}\xspace}

\def\to                 {\ensuremath{\rightarrow}\xspace}

\def\pep2{PEP-II}
\def\BF{$B$ Factory}

\def\gsim{{~\raise.15em\hbox{$>$}\kern-.85em
          \lower.35em\hbox{$\sim$}~}\xspace}
\def\lsim{{~\raise.15em\hbox{$<$}\kern-.85em
          \lower.35em\hbox{$\sim$}~}\xspace}

\def\epstag  {\ensuremath{\varepsilon_{\rm tag}}\xspace}

\def\jetset74   {\mbox{\tt Jetset \hspace{-0.5em}7.\hspace{-0.2em}4}\xspace}

\usepackage{relsize}

\def\Mnu{\ensuremath{{\cal M}^2}}

\def\Ms{\ensuremath{{\cal M}_s^2}}
\def\M{\ensuremath{{\cal M}_1^2}}
\def\MM{\ensuremath{{\cal M}_2^2}}

\def\fzz{\ensuremath{f_{00}}}
\def\Ns{\ensuremath{N_s}}
\def\Nd{\ensuremath{N_d}}

\long\def\inst#1{\par\nobreak\kern 4pt\nobreak
    {\it #1}\par\vskip 10pt plus 3pt minus 3pt}

\RequirePackage{xspace}

\usepackage{relsize}
\def\babar{\mbox{\slshape B\kern-0.1em{\smaller A}\kern-0.1em
    B\kern-0.1em{\smaller A\kern-0.2em R}}}

\def\en         {\ensuremath{e^-}\xspace}   
\def\ep         {\ensuremath{e^+}\xspace}
 
\def\epem       {\ensuremath{e^+e^-}\xspace}

\def\mup        {\ensuremath{\mu^+}\xspace}
\def\mun        {\ensuremath{\mu^-}\xspace} 
\def\mumu       {\ensuremath{\mu^+\mu^-}\xspace}

\def\taup       {\ensuremath{\tau^+}\xspace}
\def\taum       {\ensuremath{\tau^-}\xspace}

\def\ellell     {\ensuremath{\ell^+ \ell^-}\xspace}

\def\nub        {\ensuremath{\overline{\nu}}\xspace}
\def\nunub      {\ensuremath{\nu{\overline{\nu}}}\xspace}
\def\nub        {\ensuremath{\overline{\nu}}\xspace}
\def\nunub      {\ensuremath{\nu{\overline{\nu}}}\xspace}
\def\nue        {\ensuremath{\nu_e}\xspace}
\def\nueb       {\ensuremath{\nub_e}\xspace}

\def\num        {\ensuremath{\nu_\mu}\xspace}
\def\numb       {\ensuremath{\nub_\mu}\xspace}

\def\nut        {\ensuremath{\nu_\tau}\xspace}
\def\nutb       {\ensuremath{\nub_\tau}\xspace}

\def\nul        {\ensuremath{\nu_\ell}\xspace}
\def\nulb       {\ensuremath{\nub_\ell}\xspace}

\def\g     {\ensuremath{\gamma}\xspace}
\def\gaga  {\ensuremath{\gamma\gamma}\xspace}  
\def\ggstar{\ensuremath{\gamma\gamma^*}\xspace}

\def\piz   {\ensuremath{\pi^0}\xspace}

\def\ppz   {\ensuremath{\pi^0\pi^0}\xspace}
\def\pip   {\ensuremath{\pi^+}\xspace}
\def\pim   {\ensuremath{\pi^-}\xspace}
\def\pipi  {\ensuremath{\pi^+\pi^-}\xspace}

\def\pimp  {\ensuremath{\pi^\mp}\xspace}

\def\kaon  {\ensuremath{K}\xspace}
\def\Kbar  {\kern 0.2em\overline{\kern -0.2em K}{}\xspace}

\def\Kz    {\ensuremath{K^0}\xspace}
\def\Kzb   {\ensuremath{\Kbar^0}\xspace}
\def\KzKzb {\ensuremath{\Kz \kern -0.16em \Kzb}\xspace}
\def\Kp    {\ensuremath{K^+}\xspace}
\def\Km    {\ensuremath{K^-}\xspace}
\def\Kpm   {\ensuremath{K^\pm}\xspace}

\def\KpKm  {\ensuremath{\Kp \kern -0.16em \Km}\xspace}
\def\KS    {\ensuremath{K^0_{\scriptscriptstyle S}}\xspace} 
\def\KL    {\ensuremath{K^0_{\scriptscriptstyle L}}\xspace} 
\def\Kstarz  {\ensuremath{K^{*0}}\xspace}

\def\Kstar   {\ensuremath{K^*}\xspace}

\def\Kstarp  {\ensuremath{K^{*+}}\xspace}

\def\Dbar    {\kern 0.2em\overline{\kern -0.2em D}{}\xspace}

\def\Dz      {\ensuremath{D^0}\xspace}
\def\Dzb     {\ensuremath{\Dbar^0}\xspace}
\def\DzDzb   {\ensuremath{\Dz {\kern -0.16em \Dzb}}\xspace}
\def\Dp      {\ensuremath{D^+}\xspace}
\def\Dm      {\ensuremath{D^-}\xspace}

\def\DpDm    {\ensuremath{\Dp {\kern -0.16em \Dm}}\xspace}
\def\Dstar   {\ensuremath{D^*}\xspace}

\def\Dstarz  {\ensuremath{D^{*0}}\xspace}

\def\Dstarp  {\ensuremath{D^{*+}}\xspace}
\def\Dstarm  {\ensuremath{D^{*-}}\xspace}

\def\B       {\ensuremath{B}\xspace}
\def\Bbar    {\kern 0.18em\overline{\kern -0.18em B}{}\xspace}
\def\Bb      {\ensuremath{\Bbar}\xspace}
\def\BB      {\ensuremath{\B {\kern -0.16em \Bb}}\xspace}
\def\Bz      {\ensuremath{B^0}\xspace}
\def\Bzb     {\ensuremath{\Bbar^0}\xspace}
\def\BzBzb   {\ensuremath{\Bz {\kern -0.16em \Bzb}}\xspace}
\def\NBB     {\ensuremath{N_{\BB}}}

\def\Bu      {\ensuremath{B^+}\xspace}
\def\Bub     {\ensuremath{B^-}\xspace}

\def\Bpm     {\ensuremath{B^\pm}\xspace}

\def\BpBm    {\ensuremath{\Bu {\kern -0.16em \Bub}}\xspace}

\def\BorBbar    {\kern 0.18em\optbar{\kern -0.18em B}{}\xspace}
\def\DorDbar    {\kern 0.18em\optbar{\kern -0.18em D}{}\xspace}
\def\KorKbar    {\kern 0.18em\optbar{\kern -0.18em K}{}\xspace}

\def\jpsi     {\ensuremath{{J\mskip -3mu/\mskip -2mu\psi\mskip 2mu}}\xspace}

\mathchardef\Upsilon="7107
\def\Y#1S{\ensuremath{\Upsilon{(#1S)}}\xspace}

\def\FourS {\Y4S}

\mathchardef\Deltares="7101
\mathchardef\Xi="7104
\mathchardef\Lambda="7103
\mathchardef\Sigma="7106
\mathchardef\Omega="710A

\def\Deltabar{\kern 0.25em\overline{\kern -0.25em \Deltares}{}\xspace}
\def\Lbar{\kern 0.2em\overline{\kern -0.2em\Lambda\kern 0.05em}\kern-0.05em{}\xspace}
\def\Sigbar{\kern 0.2em\overline{\kern -0.2em \Sigma}{}\xspace}
\def\Xibar{\kern 0.2em\overline{\kern -0.2em \Xi}{}\xspace}
\def\Obar{\kern 0.2em\overline{\kern -0.2em \Omega}{}\xspace}
\def\Nbar{\kern 0.2em\overline{\kern -0.2em N}{}\xspace}
\def\Xb{\kern 0.2em\overline{\kern -0.2em X}{}\xspace}

\def\X {\ensuremath{X}\xspace}

\def\BR         {{\ensuremath{\cal B}\xspace}}

\def\bpsikl     {\ensuremath{\Bz \to \jpsi \KL}\xspace}

\def\Bzbtox     {\ensuremath{\Bzb \to \X}\xspace}

\def\invfb   {\ensuremath{\mbox{\,fb}^{-1}}\xspace}

\def\mus  {\ensuremath{\rm \,\mus}\xspace}

\def\mus        {\ensuremath{\,\mu{\rm s}}\xspace}    

\def\degrees{\ensuremath{^{\circ}}\xspace}

\def\to                 {\ensuremath{\rightarrow}\xspace}
\def\pep2{PEP-II}
\def\BF{$B$ Factory}

\def\gsim{{~\raise.15em\hbox{$>$}\kern-.85em
          \lower.35em\hbox{$\sim$}~}\xspace}
\def\lsim{{~\raise.15em\hbox{$<$}\kern-.85em
          \lower.35em\hbox{$\sim$}~}\xspace}

\def\epstag  {\ensuremath{\varepsilon_{\rm tag}}\xspace}

\def\jetset74   {\mbox{\tt Jetset \hspace{-0.5em}7.\hspace{-0.2em}4}\xspace}

\usepackage{relsize}

\def\Mnu{\ensuremath{{\cal M}^2}}

\def\Ms{\ensuremath{{\cal M}_s^2}}
\def\M{\ensuremath{{\cal M}_1^2}}
\def\MM{\ensuremath{{\cal M}_2^2}}

\def\fzz{\ensuremath{f_{00}}}
\def\Ns{\ensuremath{N_s}}
\def\Nd{\ensuremath{N_d}}

\begin{document}
{\pagestyle{empty}

\begin{flushright}
\babar-CONF-\BABARPubYear/\BABARConfNumber \\
SLAC-PUB-\SLACPubNumber \\
November 2004 \\
\end{flushright}

\par\vskip 5cm

\begin{center}
\Large \bf \boldmath First Measurement of the Branching Fraction of
$e^+e^- \rightarrow \BzBzb$
\end{center}
\bigskip

\begin{center}
\large Romulus Godang\footnote{E-mail: godang@phy.olemiss.edu.}\\
\vspace{0.15cm}
{\em Department of Physics and Astronomy\\ 
University of Mississippi-Oxford, University, MS 38677}\\
\vspace{0.15cm}
and\\
\vspace{0.15cm}
{\em Stanford Linear Accelerator Center, Stanford University, Stanford, CA 94309.}
\mbox{ }\\
\end{center}
\bigskip \bigskip

\begin{center}
\large
\end{center}
We report the first measurement of the absolute branching fraction
$e^+e^- \rightarrow \BzBzb$ at the $\Y4S$ resonance using data 
collected with the \babar\ detector at the \pep2\ asymmetric-energy 
$e^+e^-$ storage ring.  The analysis is performed with partial reconstruction 
of the decay $\Bzb \rightarrow D^{*+}\ell^{-} \bar{\nu}_{\ell}$,
where the presence of a signal decay is determined using only the lepton 
and the soft pion from the $D^{*}$ decay.  By reconstructing events with 
one or two signal decays we obtain a preliminary result of
$e^+e^- \rightarrow \BzBzb = 0.486 \pm 0.010(stat.) \pm 0.009(sys.)$.
Our result does not depend on branching fractions of the $\Bzb$
and the $D^{*+}$ decay chains, on the individual simulated reconstruction efficiencies, 
on the ratio of the charged and neutral $B$ meson lifetimes, or 
on the assumption of isospin symmetry.

\vspace{0.7cm}

\begin{center}
Proceedings to the DPF 2004: Annual Meeting of the Division of Particles and Fields of APS\\
26 August-31 August 2004, Riverside, CA, USA.
\end{center}

\vspace{0.7cm}

\begin{center}
Work supported in part by the U.S. Department of Energy contracts\\
DE-AC02-76SF00515 and DE-FG05-91ER40622.
\end{center}

\newpage

} 


\section{Introduction}

Isospin violation in decays of $e^+e^- \rightarrow \BB$ at 
the $\Y4S$ resonance results in a difference between the branching fractions
$\fzz \equiv {\cal B}(e^+e^- \rightarrow \BzBzb)$  
and
$f_{+-} \equiv {\cal B}(e^+e^- \rightarrow B^+ {B}^{-})$.
The experimental value of $R^{+/0} \equiv f_{+-} / \fzz$ measured by
\babar\ is $1.006 \pm 0.036 \pm 0.031$~\cite{haleh} and 
$1.10 \pm 0.06 \pm 0.05$~\cite{babar}, 
by Belle is $1.01 \pm 0.03 \pm 0.09$~\cite{belle03}, 
by CLEO is $1.058 \pm 0.084 \pm 0.136$~\cite{godang02} and 
$1.04 \pm 0.07 \pm 0.04$~\cite{silvia}. 
Theoretical predictions for $R^{+/0}$ range from 1.03 to 1.25~\cite{eichten}.
A precision measurement of $\fzz$ or $f_{+-}$ can be used
to re-normalize all $B$ meson branching fractions, eliminating the
usual assumption that $\fzz = f_{+-} = 50\%$, and will
bring us closer to an understanding of the isospin violation 
in the $\Upsilon(4S)$ decays.

This first direct measurement of $\fzz$ is based on partial reconstruction of the decay 
$\Bzb \rightarrow D^{*+} \ell^{-}\bar{\nu}_{\ell}$.\footnote{The inclusion of 
charge-conjugate states is implied throughout this paper.}
The sample of events in which at least one $\Bzb \rightarrow D^{*+} \ell^{-}
\bar{\nu}_{\ell}$ candidate decay is found is labeled as ``single-tag sample''. 
The number of signal events in such decays is
\begin{equation}
\Ns =
      2 \NBB \fzz\, \epsilon_{s} \, 
      {\cal B}(\Bzb \rightarrow D^{*+} \ell^- \bar{\nu}_{\ell}),
\label{eq:ns}
\end{equation}
where $\NBB = (88726 \pm 23)\times 10^3$ is the total number of $\BB$
events in the data sample and $\epsilon_{s}$ is the reconstruction
efficiency of the decay $\Bzb \rightarrow D^{*+} \ell^{-}
\bar{\nu}_{\ell}$.  The technique for measuring \NBB\ is 
described elsewhere~\cite{bcounting}.
The number of signal events in which two $\Bzb \rightarrow D^{*+} 
\ell^{-}\bar{\nu}_{\ell}$ candidates are found is labeled as ``double-tag sample'':
\begin{equation}
\Nd =  
      \NBB \, \fzz\, \epsilon_{d} \,
      [{\cal B}(\Bzb \rightarrow D^{*+} \ell^- \bar{\nu}_{\ell})]^2 ,  
\label{eq:nd}
\end{equation} 
where $\epsilon_{d}$ is the efficiency to reconstruct two $\Bzb \rightarrow D^{*+}
\ell^{-}\bar{\nu}_{\ell}$ decays in the same event.
Using Eq.~(\ref{eq:ns}),~(\ref{eq:nd}) and defining $C \equiv \epsilon_{d} / \epsilon_{s}^{2}$, 
\fzz\ is given by
\begin{equation}
\fzz =  
       {C N_{s}^{2} \over {4 N_{d} \NBB} }.
\label{eq:f00}
\end{equation}

\section{Dataset and Analysis Technique}

The \babar\ data sample used in this paper consists of 81.7\invfb collected 
at the $\Y4S$ resonance and 9.6\invfb collected 40\mev below the resonance.
A detailed description of the \babar\ detector is provided
elsewhere~\cite{babar_nim}.

The decays $\Bzb \rightarrow D^{*+} \ell^{-} \bar{\nu}_{\ell}$
are partially reconstructed.
This technique has been widely used~\cite{godang02,argus,franco}.
All lepton (soft pion) candidates are required to have 
momenta between 1.5\gevc and 2.5\gevc (60\mevc and 200\mevc) 
in the $\epem$ center-of-mass (CM) frame.
The neutrino invariant mass squared is calculated by
\begin{eqnarray}
\Mnu \equiv (E_{\mbox{\rm beam}}-E_{{D^*}} - 
E_{\ell})^2-({\bf{p}}_{{D^*}} + {\bf{p}}_{\ell})^2\ ,
\label{eqn:mms}
\end{eqnarray}
where $E_{\mbox{\rm beam}}$ is the beam energy and $E_{\ell}~(E_{{D^*}})$ 
and ${\bf{p}}_{\ell}~({\bf{p}}_{{D^*}})$ are the CM energy and momentum 
of the lepton (the $D^*$ meson). 

In what follows, we use the symbol $\Ms$ to denote $\Mnu$ for any
candidate in the single-tag sample.
In the double-tag sample, we randomly choose one of the two
reconstructed $\Bzb \rightarrow D^{*+} \ell^- \nu_l$ candidates 
as ``first'' and the other as ``second''. Their $\Mnu$ values are 
labeled $\M$ and $\MM$, respectively.
We define a signal region $\Mnu > -2~$GeV$^{2}/c^4$ and a sideband 
$-8 < \Mnu < -4~$GeV$^{2}/c^4$. 
We also require that the first candidate has to fall into the signal region.
This selection increases the ratio of signal to background as much as a factor of
2 in statistics compared to that without the selection~\cite{godangdpf03}.

The continuum background events are non-resonant decays of $e^{+} e^{-} 
\rightarrow \gamma^* \rightarrow q\bar q$ where $q = u, d, s, c$.
The combinatorial \BB\ background is formed from random combinations of
reconstructed leptons and soft pions.  This background can also be due to the 
low-momentum soft pions not coming from a $D^{*}$, produced by 
production correlation between $D$ mesons and their associated pions either 
the same $B$ or the other~\cite{e791}.
The peaking $\BB$ background is composed of $\Bb \to D^{*} (n\pi)
\ell \bar{\nu}_{\ell}$ decays with or without an excited charmed
resonance $D^{**}$~\cite{e961}.

\begin{figure}
\vspace*{-2.1cm}
\begin{center}
\begin{tabular}{lr} \hspace{-0.3cm} 
\includegraphics[width=9cm]{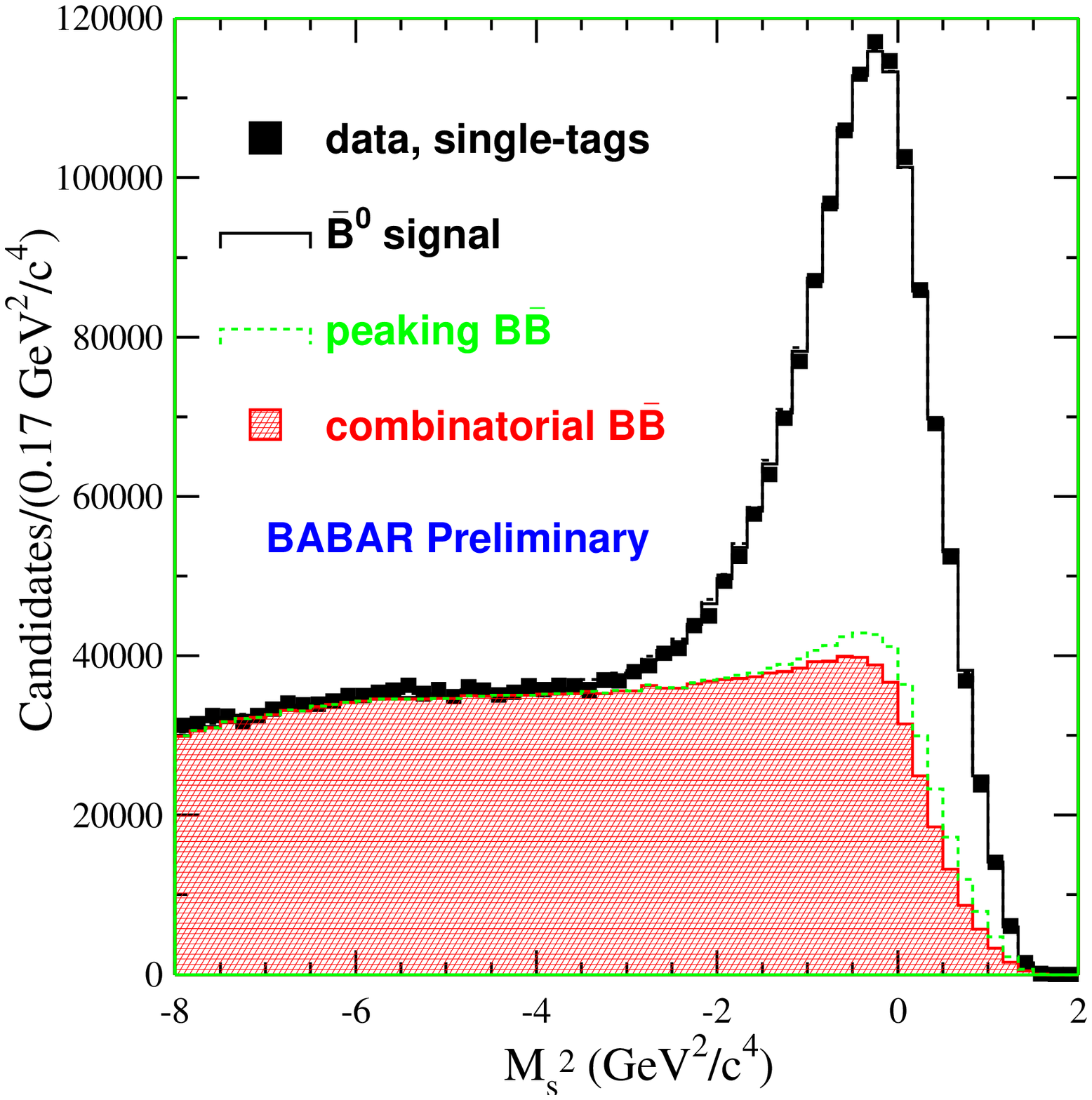}& \hspace{-1.0cm}  
\includegraphics[width=9cm]{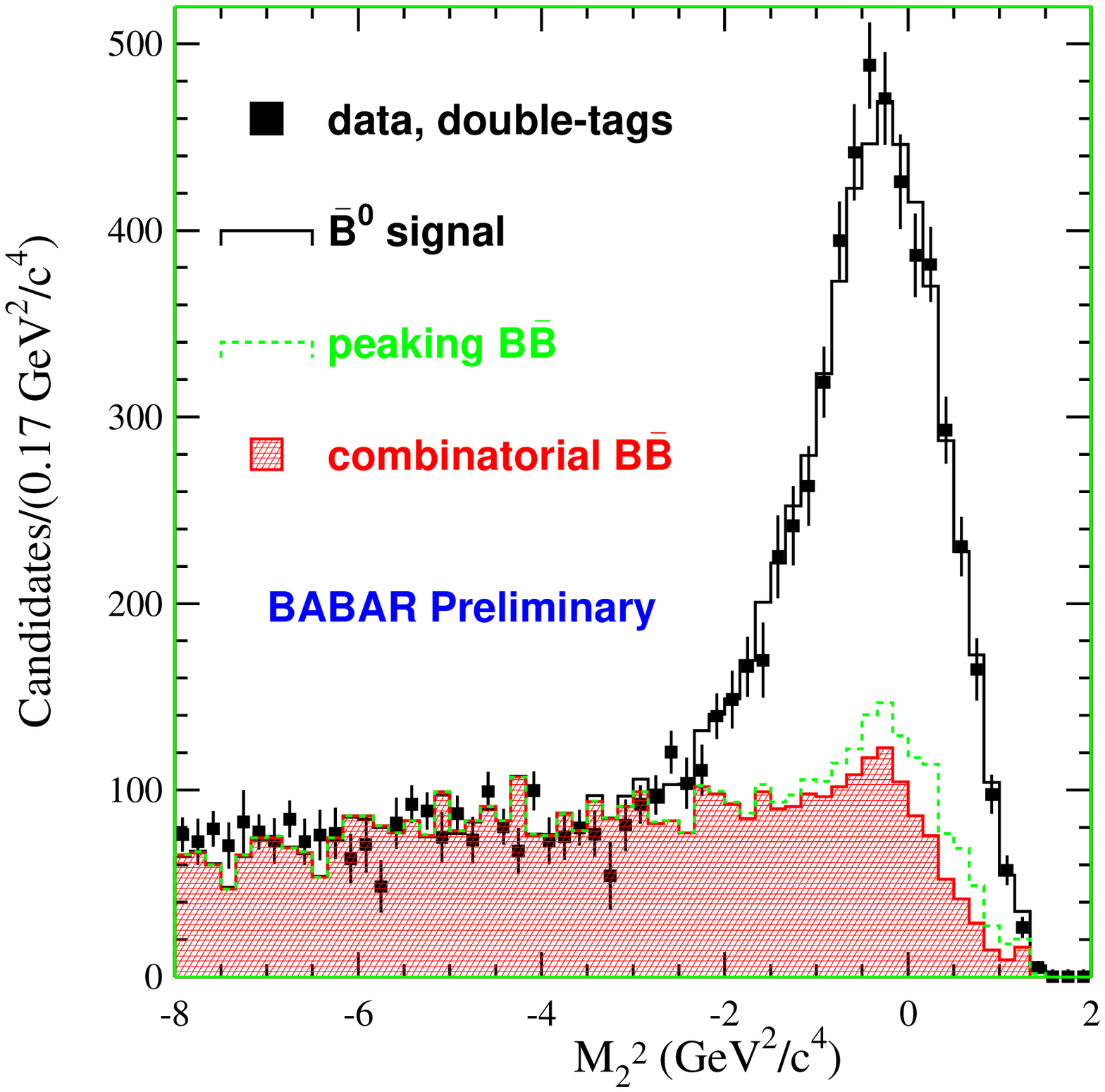} 
\end{tabular}
\vspace*{-1.0cm}
\caption{The \Ms\ (left) and \MM\ (right) distributions of the
on-resonance samples.  The continuum background has been subtracted from
the \Ms\ and \MM\ distributions. For the \MM\ distribution, the \M-combinatorial, and 
the \M-peaking have been subtracted.
The levels of the simulated signal, peaking \BB\ and
combinatorial \BB\ background contributions are obtained from the
fit.}
\label{fig:rightsign}
\end{center}
\end{figure}

The \Ms\ and \MM\ distributions are shown in Fig.~\ref{fig:rightsign}.
A $\chi^2$ binned fits yield the values $\Ns = 786300 \pm 2000$ and
$\Nd = 3560 \pm 80$. 
Using the simulation we determine $C = 0.9946 \pm 0.0078$, 
where the error is due to the finite size of the sample.

\section{Systematic Studies}

We consider several sources of systematic uncertainties in $\fzz$. 
All estimated errors are absolute systematic uncertainties in $\fzz$ 
and summarized in Table~\ref{tab:sys_error}.
\begin{table}[!htb]
\begin{center}
\caption{Summary of the absolute systematic errors for $f_{00}$.}
{\begin{tabular}{lc} \hline 
Source                                & $\delta(f_{00})$   \\ \hline
{\raisebox{-0.3ex}{\M-combinatorial}} & {\raisebox{-0.3ex}{$0.0005$}} \\
{\raisebox{-0.3ex}{\M-peaking}}       & {\raisebox{-0.3ex}{$0.0005$}} \\
Monte Carlo statistics                & $0.002$            \\
Same-charged events                   & $0.0025$           \\
$\Upsilon(4S) \to$ non-$\BB$          & $0.0025$           \\
Peaking background                    & $0.004$            \\      
Efficiency correlation                & $0.004$            \\ 
$B$-meson counting                    & $0.0055$           \\  \hline
Total                                 & $0.009$            \\  \hline 
\end{tabular}}
\label{tab:sys_error}
\end{center}
\end{table}   

\begin{enumerate}
\item The systematic uncertainty from the \M-combinatorial contribution subtraction 
in the \MM\ histogram is 0.0005.  The error is obtained by varying the total
\M-combinatorial background by its statistical error.
\item An error of 0.0005 is estimated due to the subtraction of the
\M-peaking contribution in the \MM\ histogram. 
\item An error of 0.002 is due to the finite size of the simulated sample.
\item The same-charged events lead to an error of 0.0025 on \fzz.  
\item The upper limit for the branching fraction of $\Upsilon(4S)$ decays into 
non-$\BB$ is $4\%$ at $95\%$ confidence level~\cite{nonbbar}.
The error due to such decays is 0.0025. 
\item The systematic uncertainty of the peaking background is  0.004 on \fzz. 
\item The systematic uncertainty due to the efficiency correlation is
      estimated from the Monte Carlo simulation to be 0.004.
\item The error due to the uncertainty in \NBB\ is 0.0055.
\end{enumerate}
We combine the uncertainties given above in quadrature to determine an
absolute systematic error of 0.009 in $\fzz$.
For more details see Ref.~\cite{godang04}.

\section{Summary}

In summary, we have used partial reconstruction of the decay $\Bzb \rightarrow 
D^{*+} \ell^- \nu_l$ to obtain a preliminary result of
\begin{equation}
\fzz = 0.486 \pm 0.010(stat.) \pm 0.009(sys.).
\end{equation}
This result does not depend on branching fractions of the $\Bzb$
and the $D^{*+}$ decay chains, on the individual simulated reconstruction  
efficiencies, on the ratio of the charged and neutral $B$ meson lifetimes, 
or on the assumptions of isospin symmetry.

\section{Acknowledgments}

The author would like to thank all members of the \babar\ collaboration.
This work was supported in part by the U.S. Department of Energy 
contracts DE-AC02-76SF00515 and DE-FG05-91ER40622.

\end{document}